\documentclass[10pt,conference,twocolumn]{IEEEtran}
\ifCLASSINFOpdf
\usepackage[pdftex]{graphicx}
\DeclareGraphicsExtensions{.pdf,.jpeg,.png}
\else
\usepackage[dvips]{graphicx}
\DeclareGraphicsExtensions{.eps}
\fi
\usepackage{amsmath}
\usepackage{amssymb}
\usepackage{cancel}
\usepackage{breqn}
\usepackage{multirow}
\usepackage{rotating}
\usepackage{verbatim}
\usepackage{array}
\usepackage[caption=false, font=footnotesize]{subfig}

\newcommand{\Expect}{{\rm I\kern-.5em E}}
\newtheorem{definition}{Definition}
\usepackage[utf8]{inputenc}
\usepackage[backend=biber,style=ieee,doi=false]{biblatex}
\usepackage{ulem}

\begin{document}
	\normalem
%
\title{NOMA Assisted Joint Broadcast and Multicast Transmission in 5G Networks}

\author{
	\authorblockN{Pol Henarejos, Musbah Shaat and Monica Navarro}\\
	\authorblockA{Centre Tecnol\`ogic de Telecomunicacions de Catalunya (CTTC/CERCA)\\Parc Mediterrani de la Tecnologia, Av. Carl Friedrich Gauss 7, 08860, Castelldefels, Barcelona, Spain.\\Phone: +34936452900, Fax: +34936452901\\ Email:\{pol.henarejos,musbah.shaat,monica.navarro\}@cttc.es}}

\maketitle

\begin{abstract}
In this paper, we employ the non-orthogonal multiple access (NOMA) technique to convey joint broadcast and multicast streams to  a set of users. Thanks to the spatial beamforming, different groups of users is able to receive different streams in addition to the common broadcast information. With the proposed scheme, the same time-frequency resources can be shared between different streams, without requiring additional bandwidth. The transmitter implementation is presented and two receiver classes are considered based on  Successive Interference Cancellation (SIC) and Joint Decoding (JD) approaches. In addition to the performance assessment via simulation, a real hardware proof of concept implementation of the proposed technique is performed in order to show the practical viability of the proposed scheme. 
\end{abstract}


%
\IEEEpeerreviewmaketitle

\section{Introduction}
Considering the emerging trend in user behavior and demands for multimedia services, 5G networks are predicted to support many more users per cell than current deployment. Reliable integration of broadcast and multicast services with the mobile broadband wireless networks can offer significant benefits to different services like: multimedia content delivery, file distribution, software updates, emergency messages and public warning, etc.  Current limitations of long term evolution (LTE) evolved multimedia broadcast multicast service (eMBMS), in having low Doppler speed tolerance and in limiting cyclic prefix supported length, add more challenges in supporting the users’ mobility and in increasing the service coverage. Additionally, eMBMS services do not fully exploit the spatial dimension.

Non-orthogonal transmission approaches are being considered as a promising solution to multiplex different services under common physical layer resources, increasing network throughput and efficiency \cite{7263349, 7842433}. In particular, non-orthogonal multiple access (NOMA) is a promising multiple-access scheme that exploits the differences in user channel quality (SNR) along with suitable power allocation and successive interference cancellation (SIC) at the receiver in order to improve the system throughput \cite{6966073}. In fact, NOMA applies the well-known superposition coding solution from information theory, which achieves the broadcast channel capacity for single antenna systems. However, the capacity and the capacity achieving transmission technique for the MIMO broadcast channel are in general unknown. When the channel is fixed it is known that dirty paper coding \cite{1207369} achieves capacity. However, even in this particular case the technique is far from being implemented in practice.

Multicast beamforming with superposition coding (SC) with two-stage beamforming is applied to NOMA system to support multiple users in \cite{7015589,Darsena2017}. The same scheme is used to perform multiresolution broadcasting where both high and low priority data are transmitted in the network. Layered division multiplexing (LDM) which is a form of NOMA is proposed in \cite{7842028} to simultaneously using the same resource block for multiple unicast or for broadcast transmission assuming a single frequency network (SFN). The application of NOMA to multi-user network with mixed multicasting and unicasting traffic is considered in \cite{7906532}. The beamforming and power allocation are designed in order to ensure an improvement in the unicasting performance while maintain the reception reliability of multicasting.

In this paper, we develop a novel system that is able to deliver multiple multimedia streams sharing the same frequency-time resources by enabling non-orthogonal transmissions. Moreover, thanks to spatial beamforming, we are able to sectorize such information dynamically, allowing the coexistence of different multicast and broadcast streams devoted to different group of users. Finally, we also provide a demonstration of its viability by a Proof of Concept (PoC), where the proposed techniques are implemented in real devices by means of Software Defined Radio (SDR).

The rest of this paper is organized as follows: the system model is described in the section II, where the transmitter scheme is analyzed; two different strategies of reception based on SIC and Joint Decoding (JD) are introduced in section III and IV, respectively. Finally, the PoC is described in section V.

\section{System Model}
Given a MIMO system with $N_t$ and $N_r$ antennas at transmission and reception, we aim at conveying one broadcast stream intended to all users in the system and $N_t$ multicast streams to dedicated users or to a dedicated cluster of users. Each multicast stream is conveyed using an orthonormal beam and all are superposed with the broadcast stream by adjusting the power level by the parameter $\alpha$. With this scheme, $N_t$ orthogonal beams can be constructed and therefore $2N_t$ independent streams can be conveyed. However, we also want to preserve the broadcast stream. This means that at least, one stream shall be the same for all orthogonal beams. Hence, there are $N_t+1$ available streams.

The superposed signal, which contains the contribution of the particular multicast stream in addition to the broadcast stream, is denoted as
\begin{equation}
\mathbf{x}_k=\sqrt{\alpha}\mathbf{t}_k+\sqrt{1-\alpha}\mathbf{t}_{\textrm{BC}},\ k=1...N_t,
\end{equation}
where $\mathbf{t}_k$ is the encoded block, i.e., $\mathbf{t}_k=G\left(\mathbf{u}_k\right)$ using the encoding function $G(\cdot)$, $\mathbf{t}_{\textrm{BC}}=G\left(\mathbf{u}_{\textrm{BC}}\right)$ is the encoded broadcast codeword, $\mathbf{u}_k$ and $\mathbf{u}_{\textrm{BC}}$ are the multicast and broadcast payloads, and $\alpha$ is the balancing parameter between the multicast and broadcast streams. Note that each symbol in $u_k[n]$ and $x_{\textrm{BC}}[n]$ is conveyed at $n$th time index. It is worth to mention that each encoded block can be encoded using different coding rates.

The beamforming matrix $\mathbf{B}\in\mathbb{C}^{N_t\times N_t}$ contains the beamformers in each column. The channel matrix $\mathbf{H}\in\mathbb{C}^{N_r\times N_t}$ is assumed as a random variable, with arbitrary statistic. For the particular $m$th user, the received signal at the $n$th discrete time instant is 
\begin{equation}
\mathbf{y}_m[n]=\mathbf{H}_m[n]\mathbf{B}\mathbf{x}[n]+\mathbf{w}_m[n],
\label{eq:sysmod}
\end{equation}
where $\mathbf{y}_m[n]\in\mathbb{C}^{N_r}$ is the received signal by the $m$th user, $\mathbf{H}_m\in\mathbb{C}^{N_r\times N_t}$ is the channel between the transmitter and user, $\mathbf{x}[n]=\begin{pmatrix}
x_1[n] & \cdots & x_{N_t}[n]\end{pmatrix}^T$ is the transmitted vector, and $\mathbf{w}_m\in\mathbb{C}^{N_r}$ is the Additive White Gaussian Noise (AWGN).

From now on, we assume that the user $m$ belongs to the $k$th multicast group. 

\section{Broadcast and Multicast SIC Decoding}
The system model can be envisaged as the contribution of desired signals (the broadcast and intended multicast stream), the interference of other multicast streams and the noise. Thus, expanding \eqref{eq:sysmod}, we can group the different contributions as
\begin{equation}
\begin{split}
\mathbf{y}_m[n]&=\mathbf{H}_m[n]\left(\underbrace{\sqrt{\alpha}\mathbf{b}_kt_k[n]}_{\textrm{Multicast Stream}}+\underbrace{\sqrt{1-\alpha}t_{\textrm{BC}}[n]\sum_{k=1}^{N_t}\mathbf{b}_k}_{\textrm{Broadcast Stream}}+\right.\\
&\left.\underbrace{\sqrt{\alpha}\sum_{k'\neq k}^{N_t}\mathbf{b}_{k'}t_{k'}[n]}_{\textrm{Inter-beam Interference}}\right)+\mathbf{w}_m[n]\\
&=\mathbf{H}_m[n]\left(\underbrace{\sqrt{\alpha}\mathbf{b}_kt_k[n]}_{\textrm{Multicast Stream}}+\underbrace{\sqrt{1-\alpha}\mathbf{B1}t_{\textrm{BC}}[n]}_{\textrm{Broadcast Stream}}+\right.\\
&\left.\underbrace{\sqrt{\alpha}\breve{\mathbf{B}}_k\breve{\mathbf{t}}_k[n]}_{\textrm{Inter-beam Interference}}\right)+\mathbf{w}_m[n],\\
\label{eq:sysmod2}
\end{split}
\end{equation}
where $\mathbf{1}$ is all-ones vector, $\mathbf{B}=\begin{pmatrix}
\mathbf{b}_1&\cdots&\mathbf{b}_{N_t}\end{pmatrix}^T$, $\breve{\mathbf{B}}_k$ is the beamforming matrix without the $k$th column and $\breve{\mathbf{t}}_k[n]=\begin{pmatrix}
t_1[n]&\cdots&t_{N_t}[n]\end{pmatrix}^T\setminus t_k[n]$.

To estimate the symbols $t_k[n]$ and $t_{\textrm{BC}}[n]$, the Minimum Mean Square Error (MMSE) estimator is employed with SIC. The generalized MMSE expression can be summarized as
\begin{definition}[MMSE]
	Given a linear process such that $\mathbf{y}=\mathbf{Ax}+\mathbf{z}$, where $\mathbf{A}$ is a deterministic and known matrix, $\mathbf{x}$ is an unknown zero mean vector and $\mathbf{z}$ is a zero mean uncorrelated random vector, the MMSE estimation of $\mathbf{x}$ is defined as $\hat{\mathbf{x}}=\mathbf{Wy}$ where $\mathbf{W}=\mathbf{C}_{xx}\mathbf{A}^H\left(\mathbf{C}_{zz}+\mathbf{AC}_{xx}\mathbf{A}^H\right)^{-1}$.
	\label{eq:defmmse}
\end{definition}

Depending on the knowledge of the precoding matrix at receiver, i.e., whether the user knows the full $\mathbf{B}$ or knows $\mathbf{b}_k$, different mathematical formulation can be expressed as in the sequel subsections.

\subsection{Users with full form knowledge of $\mathbf{B}$}
In this case, users only know the full form of matrix $\mathbf{B}$. It corresponds to the scenario where the beamforming matrix is static and offline-known by all the users or it is signaled to users in a semi-static basis. For the sake of clarity, we omit the $n$ index. Based on MMSE definition, the parameters as follows
\begin{equation}
\begin{split}
\mathbf{x}&\triangleq t_{\textrm{BC}}\\
\mathbf{A}&\triangleq \sqrt{1-\alpha}\mathbf{H}_m\mathbf{B1}\\
\mathbf{z}&\triangleq \sqrt{\alpha}\mathbf{H}_m\mathbf{Bt}_{\textrm{MC}}+\mathbf{w}_m\\
\mathbf{C}_{xx}&=1\\
\mathbf{C}_{zz}&=\alpha\mathbf{H}_m\mathbf{B}\Expect\left\{\mathbf{t}_{\textrm{MC}}\mathbf{t}_{\textrm{MC}}^H\right\}\mathbf{B}^H\mathbf{H}_m^H+\sigma_{\mathbf{w}_m}^2\mathbf{I}\\
&=\frac{\alpha}{N_t}\mathbf{H}_m\mathbf{H}_m^H+\sigma_{\mathbf{w}_m}^2\mathbf{I},
\end{split}
\end{equation}
where $\mathbf{t}_{\textrm{MC}}=\begin{pmatrix}
t_1&\cdots t_{N_t}\end{pmatrix}^T$ contains the $N_t$ multicast symbols and $\Expect\{\cdot\}$ is the expectation operator. To estimate the broadcast symbol, we consider multicast symbols as interference. Thus, the expression of the MMSE filter for the broadcast symbol has the following expression
\begin{equation}
\begin{split}
\mathbf{W}_{\textrm{BC}}&=\sqrt{1-\alpha}\mathbf{1}^H\mathbf{B}^H\mathbf{H}_m^H\left(\frac{\alpha}{N_t}\mathbf{H}_m\mathbf{H}_m^H\right.\\
&\left.+\sigma_{\mathbf{w}_m}^2\mathbf{I}+(1-\alpha)\mathbf{H}_m\mathbf{B11}^H\mathbf{B}^H\mathbf{H}_m^H\right)^{-1}.
\label{eq:WBC}
\end{split}
\end{equation}

Accordingly, the broadcast symbol can be estimated as
\begin{equation}
\hat{t}_{\textrm{BC}}=\mathbf{W}_{\textrm{BC}}\mathbf{y}_m
\label{eq:hattbc}
\end{equation}
To estimate the multicast symbol, first we apply the SIC principle where the estimated broadcast symbol is subtracted from the received signal. Hence,
\begin{equation}
\breve{\mathbf{y}}_m=\mathbf{y}_m-\sqrt{1-\alpha}\mathbf{H}_m\mathbf{B1}\hat{t}_{\textrm{BC}}.
\label{eq:suby}
\end{equation}
At this stage, $\breve{\mathbf{y}}_m$ only contains the multicast symbol plus the interference of other multicast symbols and noise. Hence, the MMSE definitions can be described as
\begin{equation}
\begin{split}
\mathbf{x}&\triangleq t_k\\
\mathbf{A}&\triangleq \sqrt{\alpha}\mathbf{H}_m\mathbf{b}_k\\
\mathbf{z}&\triangleq \sqrt{\alpha}\mathbf{H}_m\breve{\mathbf{B}}_k\breve{\mathbf{t}}_k+\mathbf{w}_m\\
\mathbf{C}_{xx}&=1\\
\mathbf{C}_{zz}&=\alpha\mathbf{H}_m\breve{\mathbf{B}}_k\Expect\left\{\mathbf{t}_k\mathbf{t}_k^H\right\}\breve{\mathbf{B}}_k^H\mathbf{H}_m^H+\sigma_{\mathbf{w}_m}^2\mathbf{I}\\
&=\alpha\mathbf{H}_m\breve{\mathbf{B}}_k\breve{\mathbf{B}}_k^H\mathbf{H}_m^H+\sigma_{\mathbf{w}_m}^2\mathbf{I},
\end{split}
\end{equation}
and the MMSE filter for the multicast symbol is expressed as
\begin{equation}
\begin{split}
\mathbf{W}_{\textrm{MC}}&=\sqrt{\alpha}\mathbf{b}_k^H\mathbf{H}_m^H\left(\sigma_{\mathbf{w}_m}^2\mathbf{I}+\alpha\mathbf{H}_m\breve{\mathbf{B}}_k\breve{\mathbf{B}}_k^H\mathbf{H}_m^H\right.\\
&\left.+\alpha\mathbf{H}_m\mathbf{b}_k\mathbf{b}_k^H\mathbf{H}_m^H\right)^{-1}\\
&=\sqrt{\alpha}\mathbf{b}_k^H\mathbf{H}_m^H\left(\sigma_{\mathbf{w}_m}^2\mathbf{I}+\frac{\alpha}{N_t}\mathbf{H}_m\mathbf{H}_m^H\right)^{-1}.
\label{eq:WMC}
\end{split}
\end{equation}

Finally, the multicast symbol can be estimated as
\begin{equation}
\hat{t}_k=\mathbf{W}_{\textrm{MC}}\breve{\mathbf{y}}_m.
\label{eq:hattk}
\end{equation}
Once $\hat{t}_{\textrm{BC}}$ and $\hat{t}_k$ are estimated, both are stacked to compose $\mathbf{t}_{\textrm{BC}}$ and $\mathbf{t}_k$ vectors, respectively. Thus, $k$th multicast payload and the broadcast one are decoded as
\begin{equation}
\begin{split}
\hat{\mathbf{u}}_{\textrm{BC}}=G^{-1}\left(\mathbf{t}_{\textrm{BC}}\right)\\
\hat{\mathbf{u}}_k=G^{-1}\left(\mathbf{t}_k\right),
\end{split}
\end{equation}
where $G^{-1}$ is the inverse procedure of channel coding. 

The proposed scheme accepts two refinements depending on 1) the feedback error correction motivated by the error correction property of the channel coding, and 2) using a pre-equalization stage to increase the compatibility of existing schemes.

\subsubsection{With Feedback Error Correction}
One of the advantages of using coded symbols is the fact that the channel coding can correct errors. To exploit this advantage, the receiver can stack first the estimated broadcast symbols, decode and re-encode them and finally subtract from the received signal. In this case, \eqref{eq:suby} is replaced by
\begin{equation}
\breve{\mathbf{y}}_m=\mathbf{y}_m-\sqrt{1-\alpha}\mathbf{H}_m\mathbf{B1}\bar{\mathbf{t}}_{\textrm{BC}},
\label{eq:subyfe}
\end{equation}
where $\bar{\mathbf{t}}_{\textrm{BC}}=G\left(\hat{\mathbf{u}}_{\textrm{BC}}\right)$ is the re-encoded broadcast codeword. 

Finally, broadcast and multicast symbols are estimated by using \eqref{eq:hattbc} and \eqref{eq:hattk}, where $\mathbf{y}_m$ is replaced by $\breve{\mathbf{y}}_m$ given in \eqref{eq:subyfe}. Although this scheme exploits the correction properties of channel coding, it increases the latency, since the receiver cannot perform parallel decoding of the multicast and broadcast symbols.

\subsubsection{With Pre-Equalization}
A possible approach is pre-equalizing the channel matrix by an MMSE filter or other standard equalizers of the channel matrix. That is equivalent to: 1) equalize the channel matrix, and 2) equalize precoding matrix and superposed codewords. The pre-equalized received signal with a MMSE filter is denoted as
\begin{equation}
\bar{\mathbf{y}}_m=\left(\sigma_{\mathbf{w}_m}^2\mathbf{I}+\mathbf{H}_m^H\mathbf{H}_m\right)^{-1}\mathbf{H}_m^H\mathbf{y}_m.
\label{eq:ypre}
\end{equation}

To compute the multicast and broadcast symbols, we use \eqref{eq:hattbc} and \eqref{eq:hattk} after replacing $\mathbf{y}_m$ by $\bar{\mathbf{y}}_m$ given in \eqref{eq:ypre}. Note that pre-equalization is compatible with Feedback Error Correction. Thus, \eqref{eq:suby} and \eqref{eq:subyfe} can also be used after replacing $\mathbf{y}_m$ by $\bar{\mathbf{y}}_m$ given in \eqref{eq:ypre}.

At this stage, we assume that the channel contribution in \eqref{eq:ypre} is compensated and therefore, the MMSE broadcast and multicast filters can be computed by suppressing the channel contribution in \eqref{eq:WBC} and \eqref{eq:WMC}. Thus,
\begin{equation}
\begin{split}
\mathbf{W}_{\textrm{BC}}&=\sqrt{1-\alpha}\mathbf{1}^H\mathbf{B}^H\left(\frac{\alpha}{N_t}\mathbf{I}\right.\\
&\left.+\mathbf{R}_{\tilde{\mathbf{w}}_m}+(1-\alpha)\mathbf{B11}^H\mathbf{B}^H\right)^{-1}\\
\mathbf{W}_{\textrm{MC}}&=\sqrt{\alpha}\mathbf{b}_k^H\left(\mathbf{R}_{\tilde{\mathbf{w}}_m}+\frac{\alpha}{N_t}\mathbf{I}\right)^{-1},
\end{split}
\end{equation}
where
\begin{equation}
\begin{split}
\tilde{\mathbf{w}}_m&=\left(\sigma_{\mathbf{w}_m}^2\mathbf{I}+\mathbf{H}_m^H\mathbf{H}_m\right)^{-1}\mathbf{H}_m^H\mathbf{w}_m\\
\mathbf{R}_{\tilde{\mathbf{w}}_m}&=\Expect\left\{\tilde{\mathbf{w}}_m\tilde{\mathbf{w}}_m^H\right\}\\
&=\sigma_{\mathbf{w}_m}\mathbf{H}_m^H\mathbf{H}_m\left(\sigma_{\mathbf{w}_m}^2\mathbf{I}+\mathbf{H}_m^H\mathbf{H}_m\right)^{-2}.
\end{split}
\end{equation}

On one side, although this scheme simplifies the expressions, it propagates the errors produced by the pre-equalization and colourizes the noise autocorrelation matrix. On the other side, it increases the compatibility with the previous deployments, since the receiver can be upgraded without modifying the existing signal processing blocks.

\subsection{Users with beam weight knowledge}
This is the case where the users only know their beam weights. Users estimate beam weights based on pilots and are not able to estimate the adjacent beam weights since they are almost orthogonal to their channel. This corresponds to the scenario where the transmitter optimizes the precoding matrix dynamically while the users have to estimate their respective beam weights.

Since $\mathbf{B}$ is partially known (indeed, only $\mathbf{b}_k$ is known), it means that $\breve{\mathbf{B}}_k$ is also unknown and the expressions have to be rewritten based on this constraint. Accordingly, we assume that the channel matches with the intended beam and not matches with the rest of the beams. In other words, we assume that
\begin{equation}
\|\mathbf{H}_m\mathbf{b}_k\|\gg\|\mathbf{H}_m\mathbf{b}_j\|,\ k\neq j.
\end{equation}
This assumption approximates \eqref{eq:sysmod2} by taking only the contributions of $\mathbf{b}_k$ into account. Hence,
\begin{equation}
\mathbf{y}_m[n]\approx\mathbf{H}_m[n]\left(\underbrace{\sqrt{\alpha}\mathbf{b}_kt_k[n]}_{\textrm{Multicast} k\textrm{th Stream}}+\underbrace{\sqrt{1-\alpha}\mathbf{b}_kt_{\textrm{BC}}[n]}_{\textrm{Broadcast Stream}}\right)+\mathbf{w}_m[n]
\end{equation}
This is valid in the case where the channel matches with the beam $k$ and is orthogonal with the rest. Hence, the broadcast filter definitions become
\begin{equation}
\begin{split}
\mathbf{x}&\triangleq t_{\textrm{BC}}\\
\mathbf{A}&\triangleq \sqrt{1-\alpha}\mathbf{H}_m\mathbf{b}_k\\
\mathbf{z}&\triangleq \sqrt{\alpha}\mathbf{H}_m\mathbf{b}_kt_k+\mathbf{w}_m\\
\mathbf{C}_{xx}&=1\\
\mathbf{C}_{zz}&=\alpha\mathbf{H}_m\mathbf{b}_k\mathbf{b}_k^H\mathbf{H}_m^H+\sigma_{\mathbf{w}_m}^2\mathbf{I}.
\end{split}
\end{equation}

Thus, the broadcast MMSE filter is expressed as
\begin{equation}
\begin{split}
\mathbf{W}_{\textrm{BC}}&=\sqrt{1-\alpha}\mathbf{b}_k^H\mathbf{H}_m^H\left(\sigma_{\mathbf{w}_m}^2\mathbf{I}+\alpha\mathbf{H}_m\mathbf{b}_k\mathbf{b}_k^H\mathbf{H}_m^H\right.\\
&\left.+(1-\alpha)\mathbf{H}_m\mathbf{b}_k\mathbf{b}_k^H\mathbf{H}_m^H\right)^{-1}\\
&=\sqrt{1-\alpha}\mathbf{b}_k^H\mathbf{H}_m^H\left(\sigma_{\mathbf{w}_m}^2\mathbf{I}+\mathbf{H}_m\mathbf{b}_k\mathbf{b}_k^H\mathbf{H}_m^H\right)^{-1}.
\label{eq:WBCpb}
\end{split}
\end{equation}
The subtracted received signal is obtained by reconstructing by with the transmitted signal. Hence,
\begin{equation}
\breve{\mathbf{y}}_m=\mathbf{y}_m-\alpha\mathbf{H}_m\mathbf{b}_k\bar{\mathbf{t}}_{\textrm{BC}}
\end{equation}

Finally, the multicast MMSE filter is expressed as
\begin{equation}
\mathbf{W}_{\textrm{MC}}=\sqrt{\alpha}\mathbf{b}_k^H\mathbf{H}_m^H\left(\sigma_{\mathbf{w}_m}^2\mathbf{I}+\alpha\mathbf{H}_m\mathbf{b}_k\mathbf{b}_k^H\mathbf{H}_m^H\right)^{-1}.
\label{eq:WMCpb}
\end{equation}
Pre-equalization can also be applied in this case, where \eqref{eq:WBCpb} and \eqref{eq:WMCpb} are reduced to
\begin{equation}
\begin{split}
\mathbf{W}_{\textrm{BC}}&=\sqrt{1-\alpha}\mathbf{b}_k^H\left(\mathbf{R}_{\tilde{\mathbf{w}}_m}+\mathbf{b}_k\mathbf{b}_k^H\right)^{-1}\\
\mathbf{W}_{\textrm{MC}}&=\sqrt{\alpha}\mathbf{b}_k^H\left(\mathbf{R}_{\tilde{\mathbf{w}}_m}+\alpha\mathbf{b}_k\mathbf{b}_k^H\right)^{-1}.
\end{split}
\end{equation}

In summary, there are several possibilities and combinations depending on the knowledge of the beamforming matrix, on the feedback error correction and the pre-equalization. In detail:
\begin{itemize}
	\item Users know $\mathbf{B}$ fully or partially:
	\begin{itemize}
		\item Pros: the accuracy produces better results.
		\item Cons: increase the complexity and requires signalling of $\mathbf{B}$.
	\end{itemize}
\item Feedback correction is applied or not:
\begin{itemize}
	\item Pros: reconstructed signals are more precise since some errors are corrected.
	\item Cons: increase the delay since the multicast signal has to be decoded initially.
\end{itemize}
\item Pre-equalization is applied or not:
\begin{itemize}
	\item Pros: increase the accuracy and simplify the expressions. It can be used with legacy channel estimators and equalizers.
	\item Cons: propagates channel estimation errors.
\end{itemize}
\end{itemize}

\section{Joint Decoding}
In the previous schemes, the SIC principle is applied where the broadcast symbol is decoded first, reconstructed, subtracted from the received signal and then, the multicast symbol is estimated. SIC involves the error forwarding drawback, which means that if the broadcast symbol is not decoded successfully, the multicast symbol will not.
Fortunately, \eqref{eq:sysmod} can be rewritten with a joint matrix notation as
\begin{equation}
\begin{split}
\mathbf{y}_m[n]&=\mathbf{H}_m[n]\left(\begin{pmatrix}\mathbf{b}_k&\mathbf{B1}\end{pmatrix}\begin{pmatrix}\sqrt{\alpha}&0\\0&\sqrt{1-\alpha}\end{pmatrix}\begin{pmatrix}t_k[n]\\t_{\textrm{BC}}[n]\end{pmatrix}\right.\\
&\left.+\sqrt{\alpha}\breve{\mathbf{B}}_k\breve{\mathbf{t}}_k[n]\right)+\mathbf{w}_m[n]\\
&=\mathbf{H}_m[n]\left(\ddot{\mathbf{B}}_k\boldsymbol{\Delta}\ddot{\mathbf{t}}_k[n]+\sqrt{\alpha}\breve{\mathbf{B}}_k\breve{\mathbf{t}}_k[n]\right)+\mathbf{w}_m[n],
\end{split}
\end{equation}
where $\ddot{\mathbf{B}}_k=\begin{pmatrix}\mathbf{b}_k&\mathbf{B1}\end{pmatrix}$, $\boldsymbol{\Delta}=\begin{pmatrix}\sqrt{\alpha}&0\\0&\sqrt{1-\alpha}\end{pmatrix}$, and $\ddot{\mathbf{t}}_k[n]=\begin{pmatrix}t_k[n]&t_{\textrm{BC}}[n]\end{pmatrix}^T$.

With this notation, the receiver only applies one MMSE filter at once. Thus, the joint MMSE filter definitions are 
\begin{equation}
\begin{split}
\mathbf{x}&\triangleq \ddot{\mathbf{t}}_k[n]\\
\mathbf{A}&\triangleq \mathbf{H}_m\ddot{\mathbf{B}}_k\boldsymbol{\Delta}\\
\mathbf{z}&\triangleq \sqrt{\alpha}\mathbf{H}_m\breve{\mathbf{B}}_k\breve{\mathbf{t}}_k[n]+\mathbf{w}_m\\
\mathbf{C}_{xx}&=\mathbf{I}\\
\mathbf{C}_{zz}&=\alpha\mathbf{H}_m\breve{\mathbf{B}}_k\Expect\left\{\breve{\mathbf{t}}_k\breve{\mathbf{t}}_k^H\right\}\breve{\mathbf{B}}_k^H\mathbf{H}_m^H+\sigma_{\mathbf{w}_m}^2\mathbf{I}\\
&=\alpha\mathbf{H}_m\breve{\mathbf{B}}_k\breve{\mathbf{B}}_k^H\mathbf{H}_m^H+\sigma_{\mathbf{w}_m}^2\mathbf{I},
\end{split}
\end{equation}
and thus, the joint MMSE filter is expressed as
\begin{equation}
\begin{split}
\mathbf{W}_{\textrm{J}}&=\boldsymbol{\Delta}\ddot{\mathbf{B}}_k^H\mathbf{H}_m^H\left(\sigma_{\mathbf{w}_m}^2\mathbf{I}+\alpha\mathbf{H}_m\breve{\mathbf{B}}_k\breve{\mathbf{B}}_k^H\mathbf{H}_m^H\right.\\
&\left.+\mathbf{H}_m\ddot{\mathbf{B}}_k\boldsymbol{\Delta}^2\ddot{\mathbf{B}}_k^H\mathbf{H}_m^H\right)^{-1}.
\end{split}
\label{eq:WJ}
\end{equation}
Therefore, the joint payload $\ddot{\mathbf{t}}_k$ is estimated by
\begin{equation}
\ddot{\mathbf{t}}_k=\mathbf{W}_{\textrm{J}}\mathbf{y}_m.
\end{equation}

Finally, in the case where pre-equalization is performed, \eqref{eq:WJ} is reduced to
\begin{equation}
\mathbf{W}_{\textrm{J}}=\boldsymbol{\Delta}\ddot{\mathbf{B}}_k^H\left(\mathbf{R}_{\tilde{\mathbf{w}}_m}+\alpha\breve{\mathbf{B}}_k\breve{\mathbf{B}}_k^H+\ddot{\mathbf{B}}_k\boldsymbol{\Delta}^2\ddot{\mathbf{B}}_k^H\right)^{-1}.
\end{equation}

\section{Results}
In this section we describe the results obtained with joint broadcast and multicast transmissions. We simulate a scenario with a single base station and several user equipment placed at different positions within the cell. Modulation and Coding Schemes (MCS) and framing is the same as used in LTE. The simulation parameters are summarized in Table \ref{tab:radcom}.
\begin{table}[!ht]
	\centering
	\caption{Common Radio Parameters}
	\begin{tabular}{|>{\bf}c|c|}
		\hline
		Bandwidth & $5$ MHz\\\hline
		Resource Blocks & $25$\\\hline
		Occupied Resource Elements & $100$\%\\\hline
		Antennas at Base Station & $4$\\\hline
		Height of BS & $15$ meters\\\hline
		Power & $50$ dBm\\\hline
		Antennas at User Equipment & $4$\\\hline
		Number of users & $100$\\\hline
		Carrier Frequency & $1.9$ GHz\\\hline
		Cell Radius & $500$ m\\\hline
		Coding Rate & $1/3$\\\hline
		Antenna Correlation	& Low\\\hline
		Path Loss Model	& Macro cell urban area [TS 36.942]\\\hline
		$N_0$ & $-174$ dBm/Hz\\\hline
		Noise Figure & $7$ dB\\\hline
		Channel Model & ITU Pedestrian B\\\hline
		Max. Doppler Frequency & $0$ Hz\\\hline
		Outage Probability & $10^{-2}$\\\hline
	\end{tabular}
	\label{tab:radcom}
\end{table}

We analyze the coverage using SIC and JD approaches. To evaluate the performance of the proposed schemes, we employ the coverage and joint coverage metrics. The coverage metric is defined as
\begin{equation}
\textrm{Coverage}(k)=1-\frac{N'(k)}{N(k)},
\end{equation}
where $k$ corresponds to the index of the stream, $N'(k)$ are the number of users in outage ($\textrm{PER} > 0.01$), and $N(k)$ is the total number of users served by stream $k$. The joint coverage metric combines the coverage of multicast stream $k$ and the broadcast stream, ensuring that both achieve at least the same coverage individually. For example, a joint coverage of $98\%$ ensures that the multicast and broadcast streams are received with a coverage of $98\%$ each.

\begin{figure}[!ht]
	\centering
	\includegraphics[width=0.9\linewidth,clip=true]{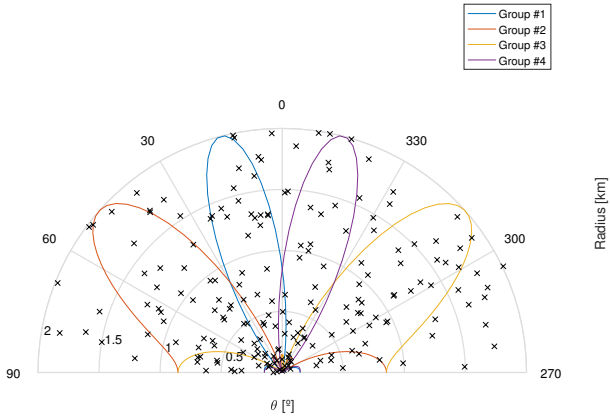}
	\caption{Distribution of user and beam pattern.}
	\label{fig:BeamPattern}
\end{figure}

\begin{figure}[!ht]
	\centering
	\subfloat[PDF]{\includegraphics[width=0.49\linewidth,clip=true]{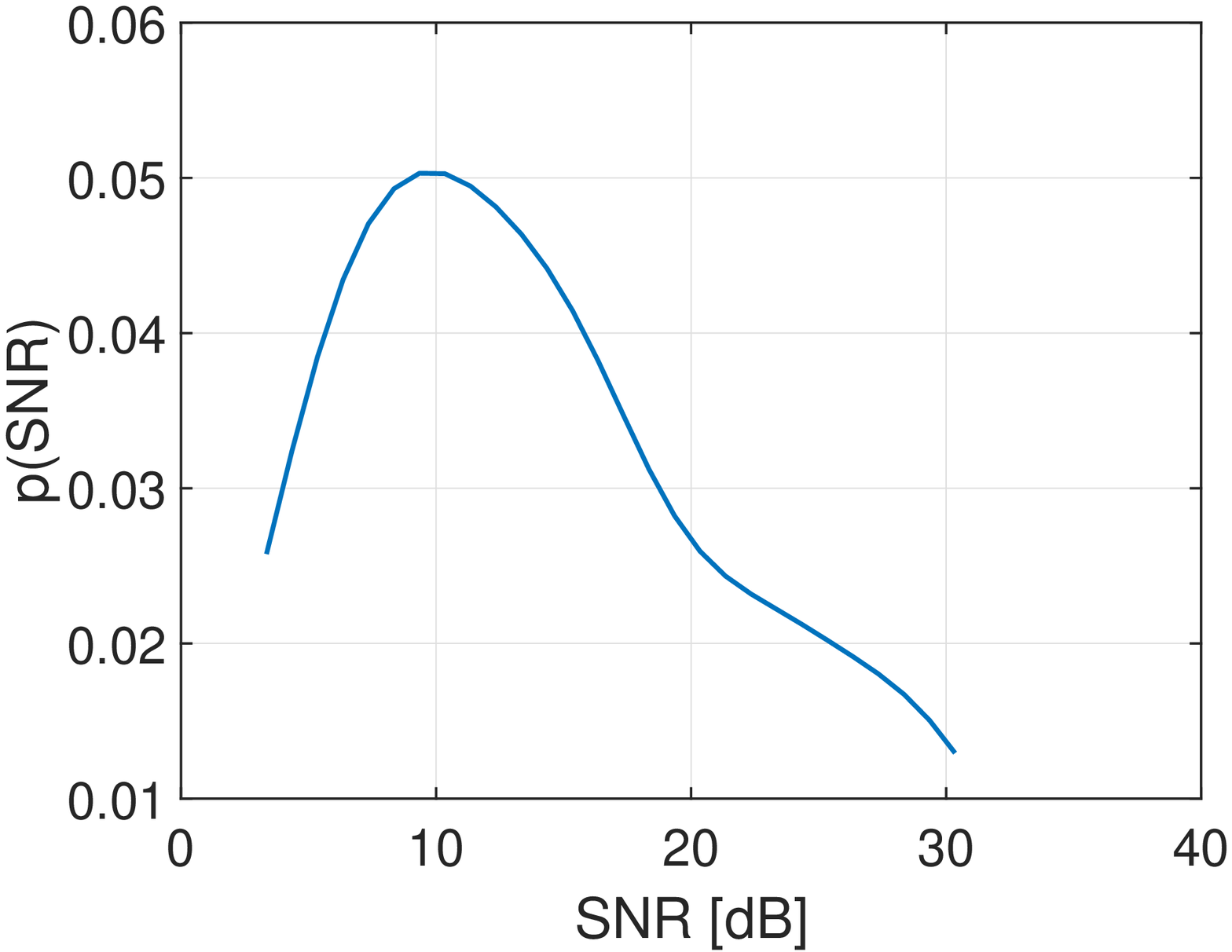}\label{fig:SNR_PDF}}
	\subfloat[CDF]{\includegraphics[width=0.49\linewidth,clip=true]{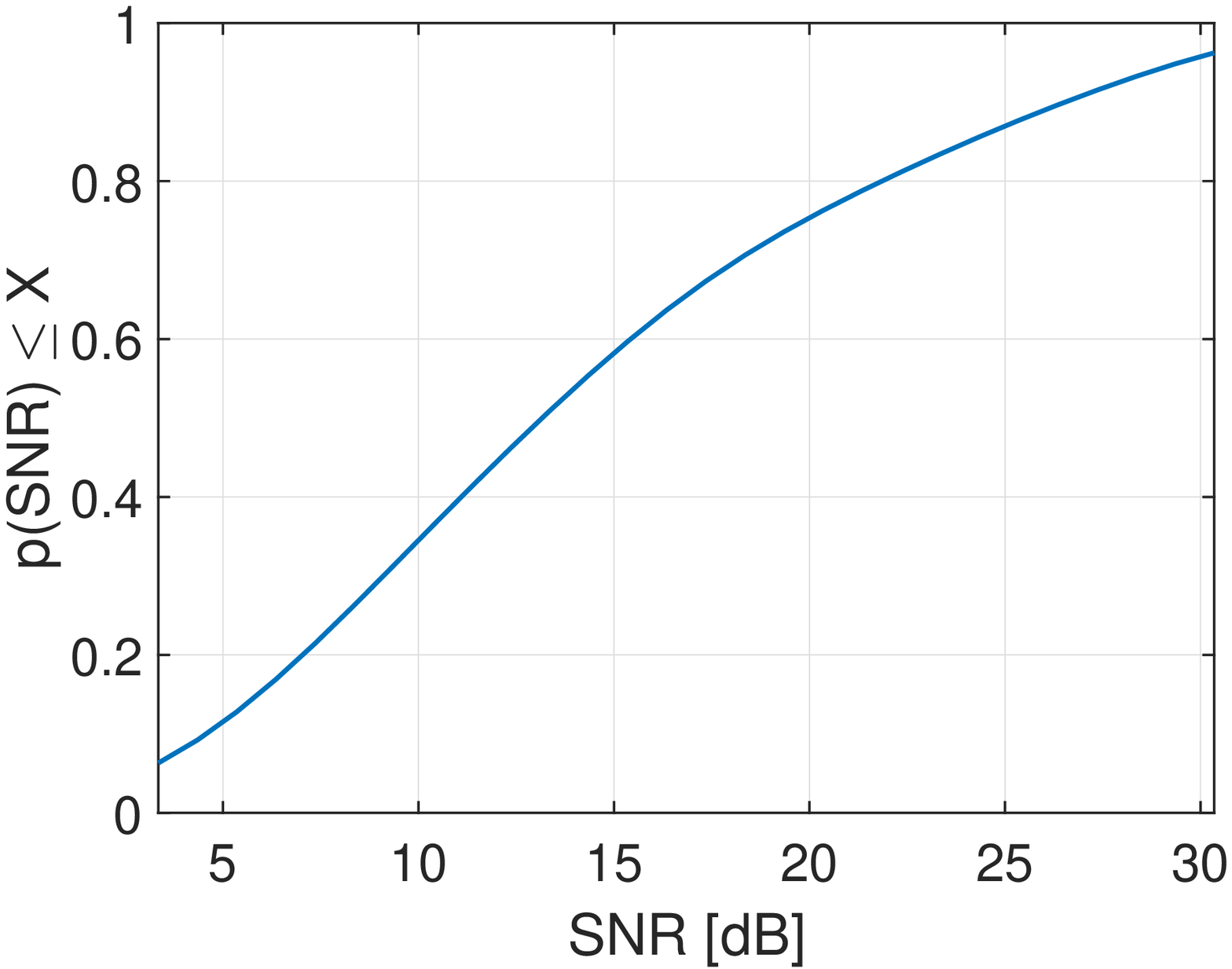}\label{fig:SNR_CDF}}
	\caption{Probability and Cumulative Density Functions of SNR.}
	\label{fig:PDF_CDF_SNR}
\end{figure}

In the considered scenario, the users are placed randomly in a cell with different distances. The path loss is different for each user and also its Signal to Noise Ratio (SNR). Fig. \ref{fig:BeamPattern} illustrates the users across the cell and the beam pattern used to convey the multicast streams jointly with the broadcast stream. Note that each group $k$ conveys the $k$th multicast stream and the broadcast stream. Fig. \ref{fig:SNR_PDF} and Fig. \ref{fig:SNR_CDF} depict the PDF and CDF of the SNR distribution in this scenario, respectively. From the SNR distribution, we can appreciate that the average SNR value is $10$ dB, with a maximum value of $30$ dB and a minimum value of $4$ dB.

\subsection{First Approach: SIC Receiver}
\begin{figure}[!ht]
	\centering
	\subfloat[Group $1$]{\includegraphics[width=0.49\linewidth,clip=true]{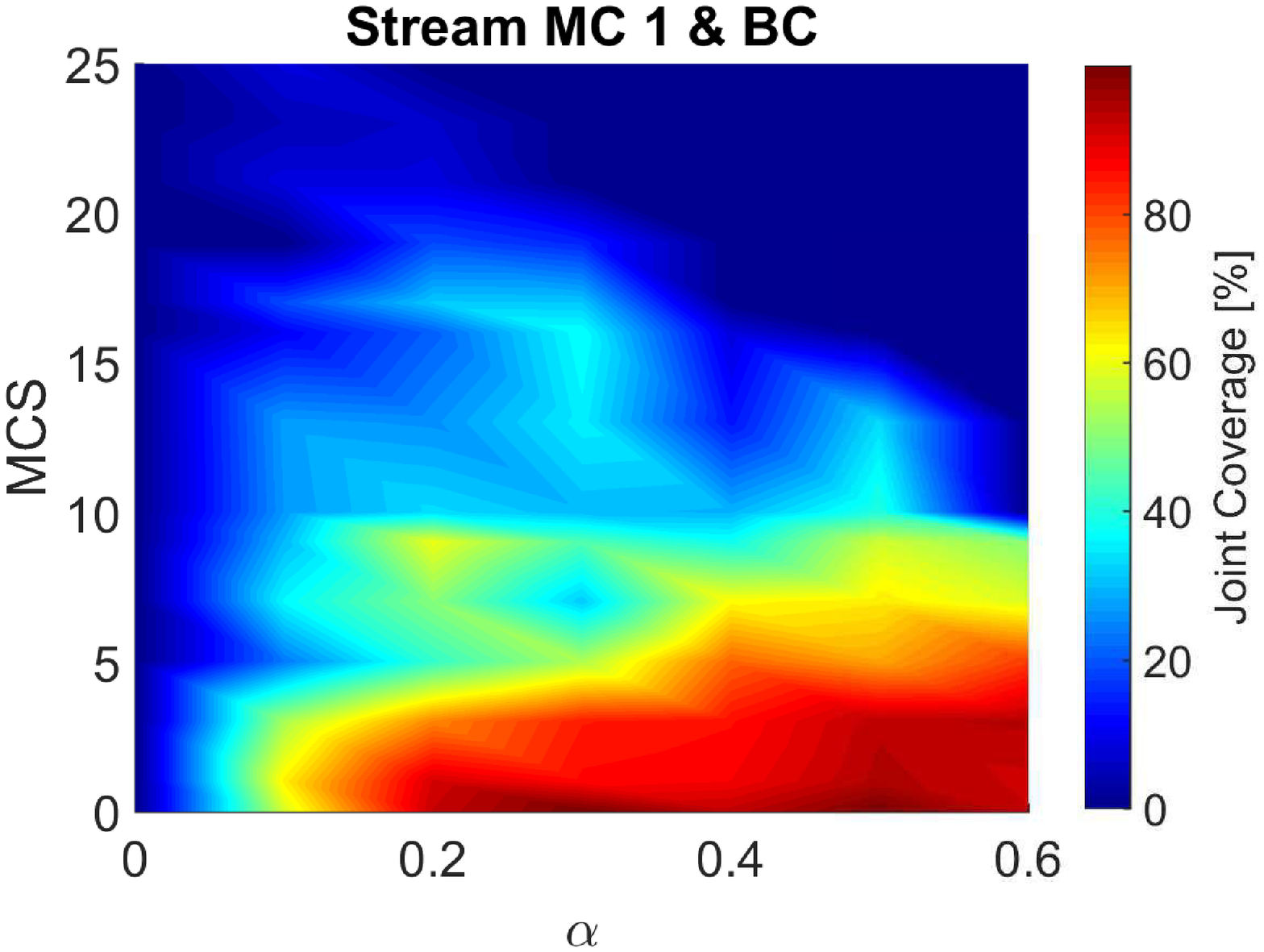}}\hfill
	\subfloat[Group $2$]{\includegraphics[width=0.49\linewidth,clip=true]{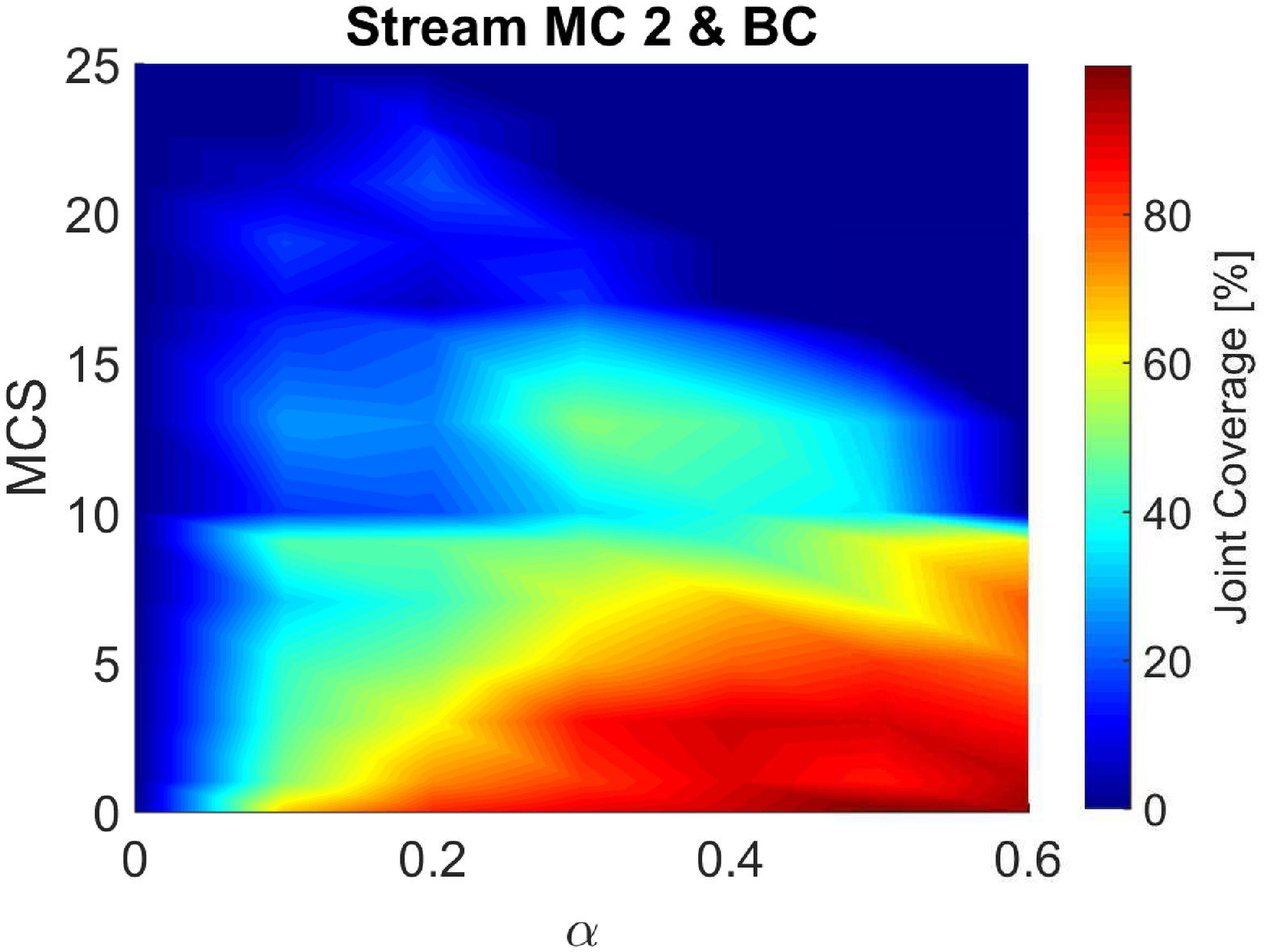}}\hfill
	\subfloat[Group $3$]{\includegraphics[width=0.49\linewidth,clip=true]{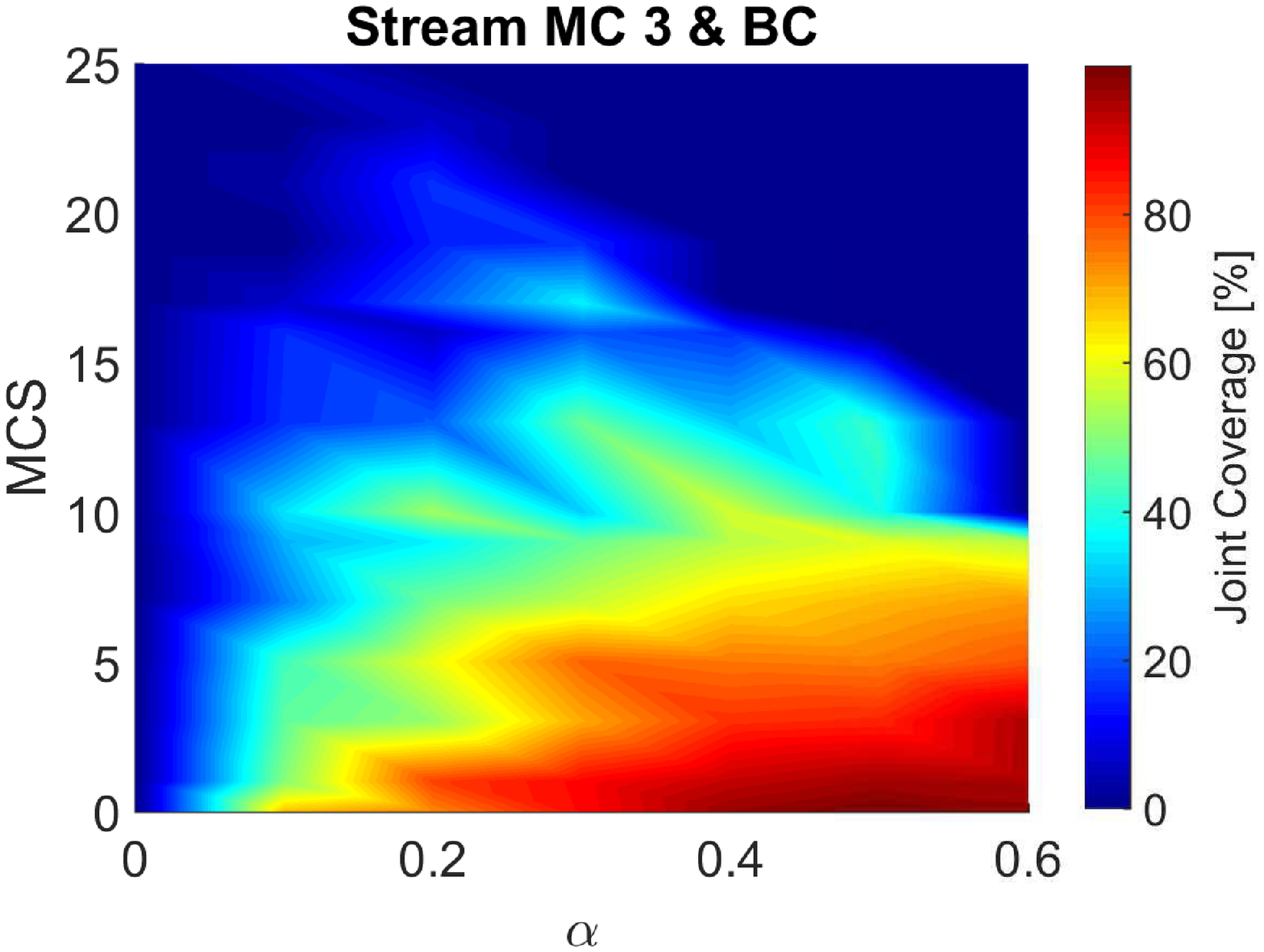}}\hfill
	\subfloat[Group $4$]{\includegraphics[width=0.49\linewidth,clip=true]{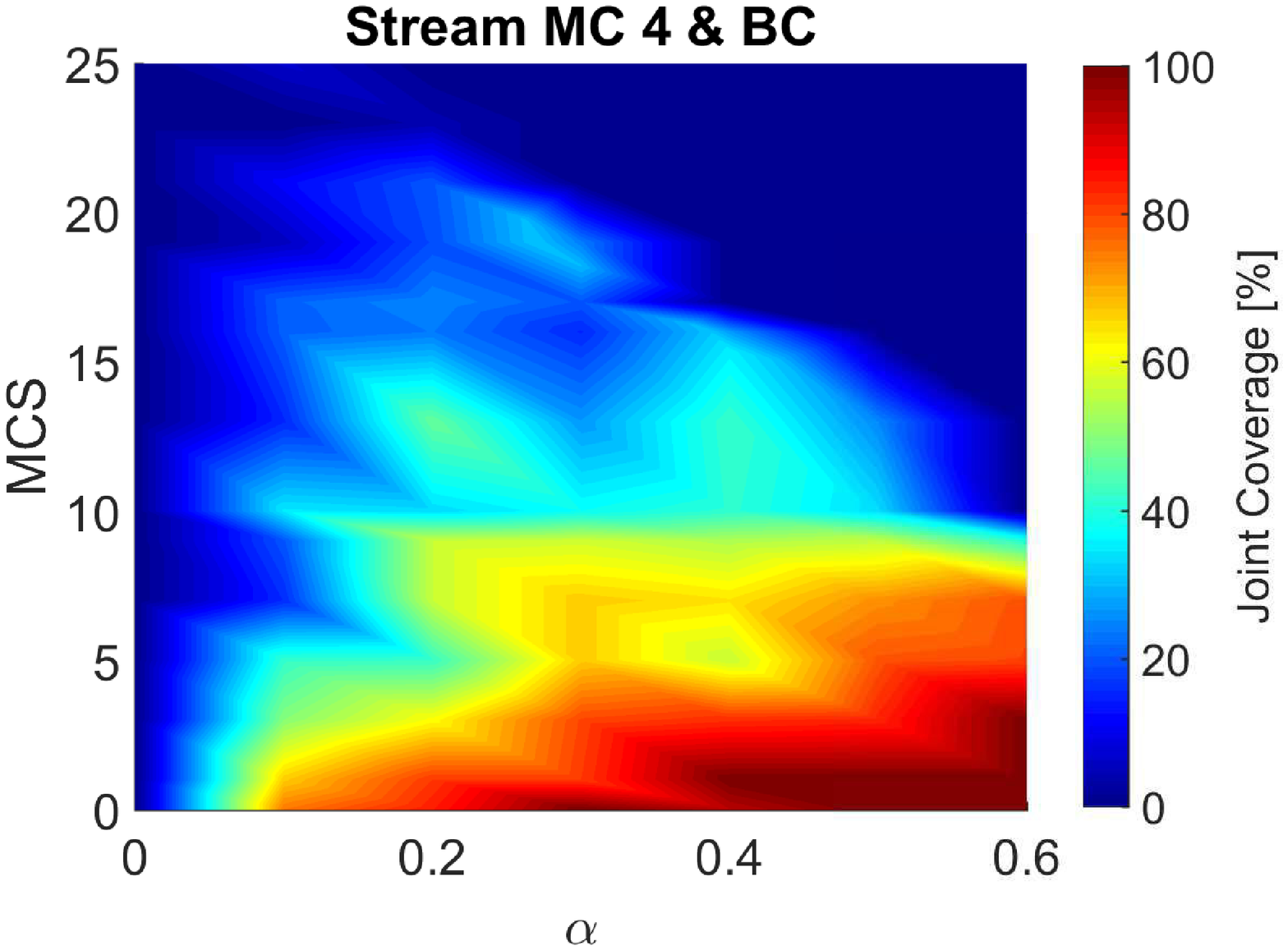}}\hfill
	\caption{Joint Coverage for the SIC Receiver.}
	\label{fig:JointCoverageSIC}
\end{figure}

The transmitter implements the SIC principle, with feedback error correction --symbols are decoded, re-encoded and subtracted--, ledge of $\mathbf{B}$, and pre-equalization of the channel matrix. Fig. \ref{fig:JointCoverageSIC} illustrates the joint coverage metric (in $\%$) of each group as a function of MCS and $\alpha$. It is clear that the maximum joint coverage is achieved for low MCS and for an $\alpha$ in the vicinity of $0.5$.

\subsection{Second Approach: Joint Decoding Receiver}
\begin{figure}[!ht]
	\centering
	\subfloat[Group $1$]{\includegraphics[width=0.49\linewidth,clip=true]{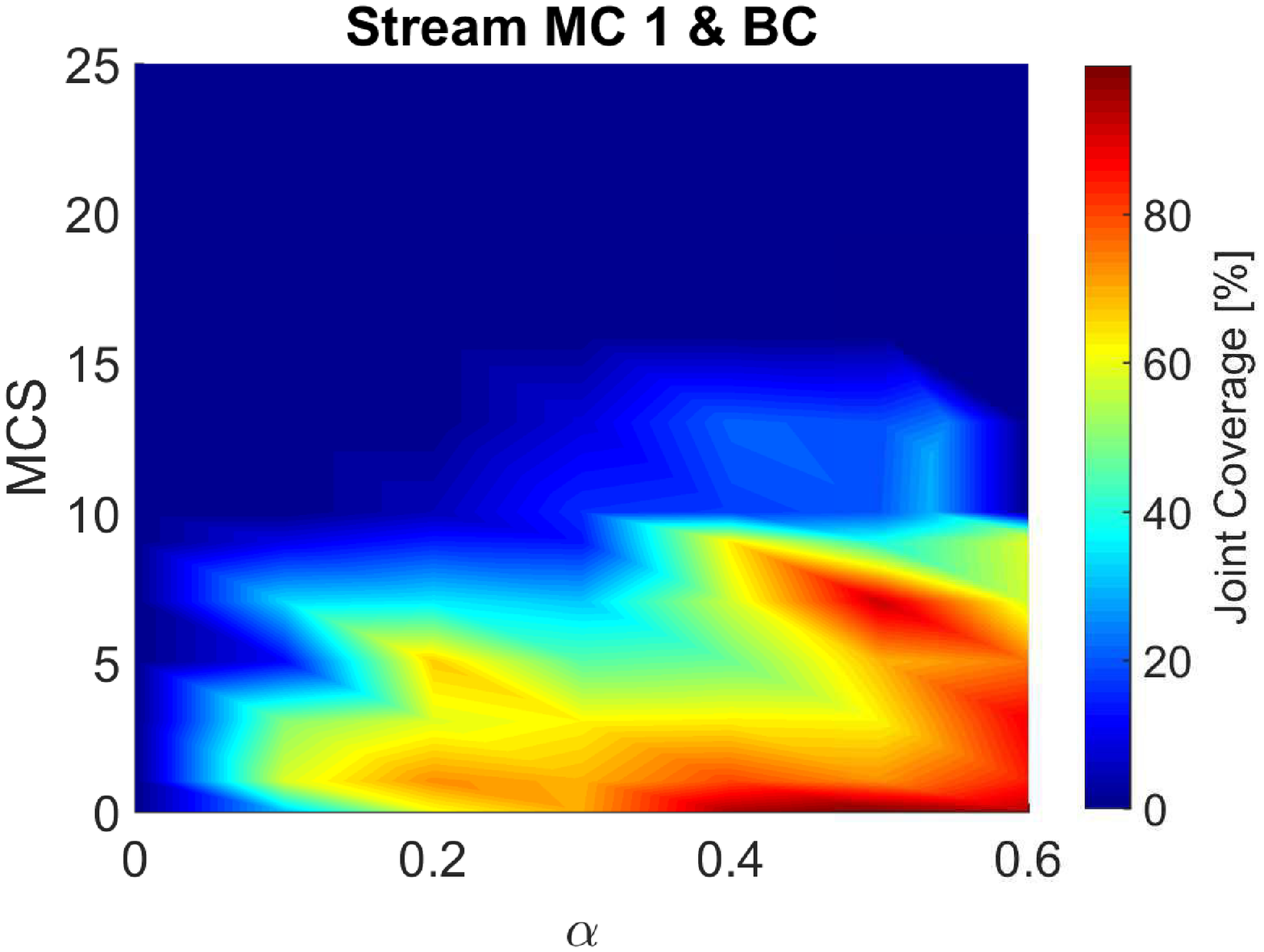}}\hfill
	\subfloat[Group $2$]{\includegraphics[width=0.49\linewidth,clip=true]{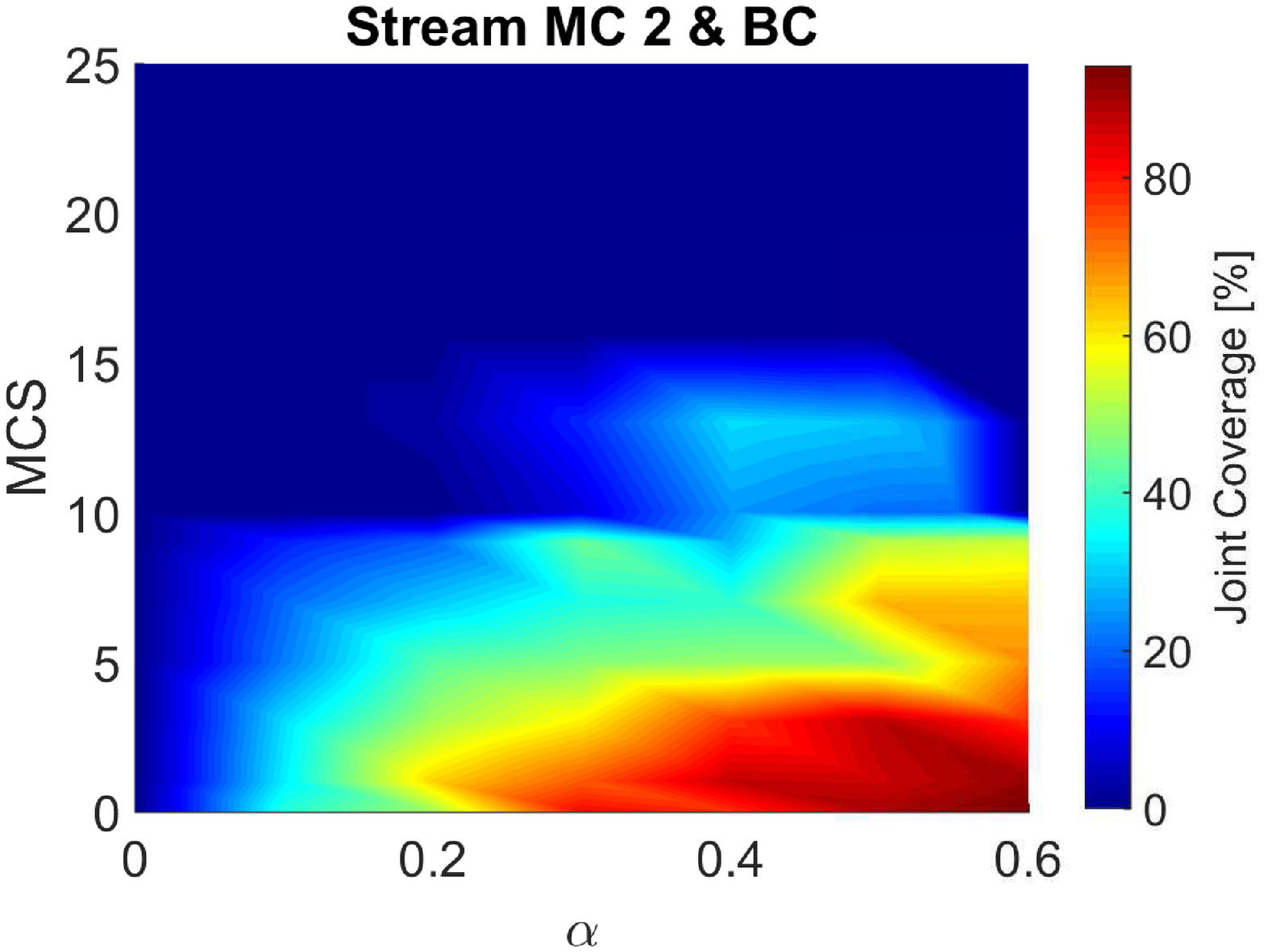}}\hfill
	\subfloat[Group $3$]{\includegraphics[width=0.49\linewidth,clip=true]{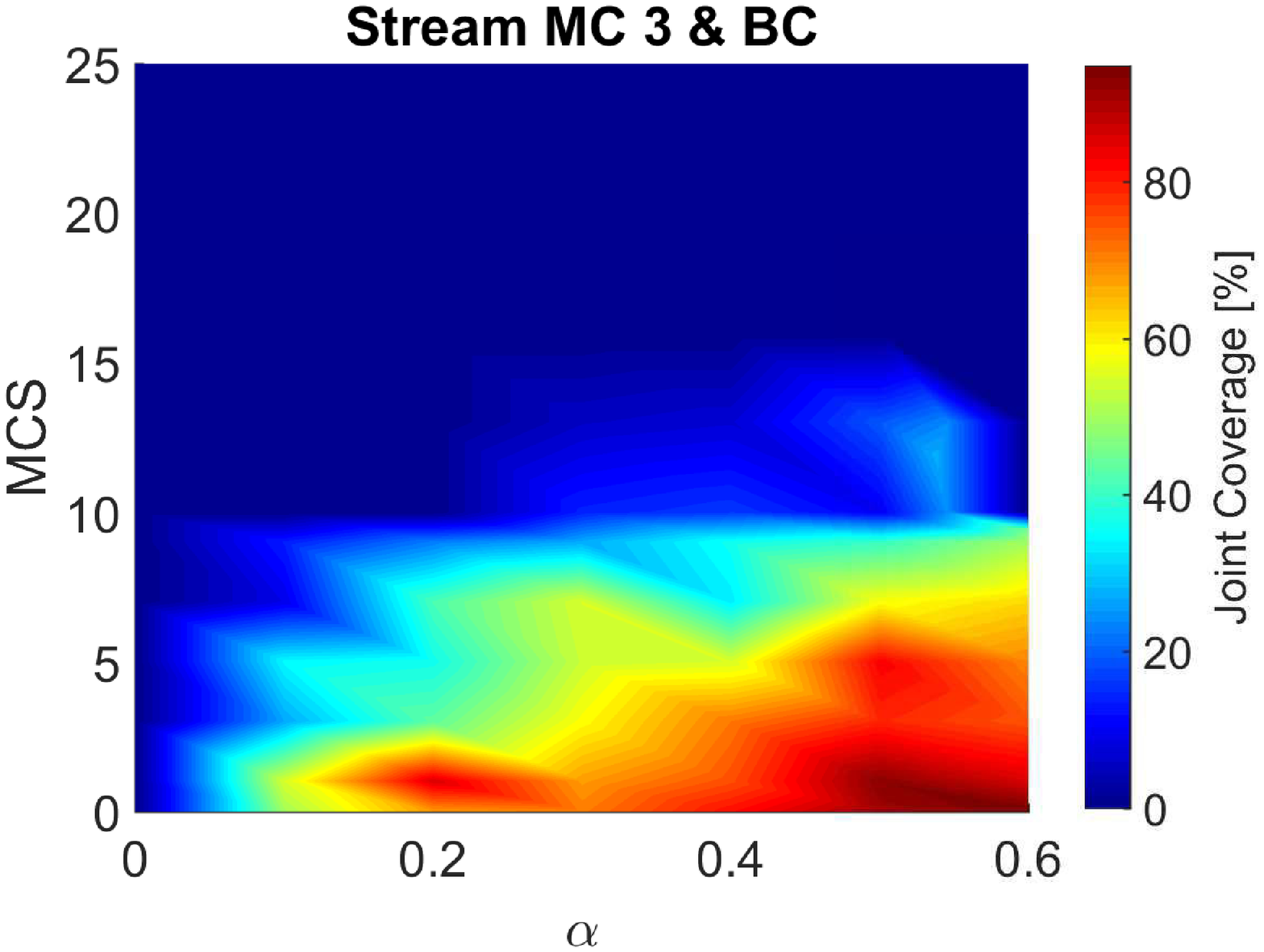}}\hfill
	\subfloat[Group $4$]{\includegraphics[width=0.49\linewidth,clip=true]{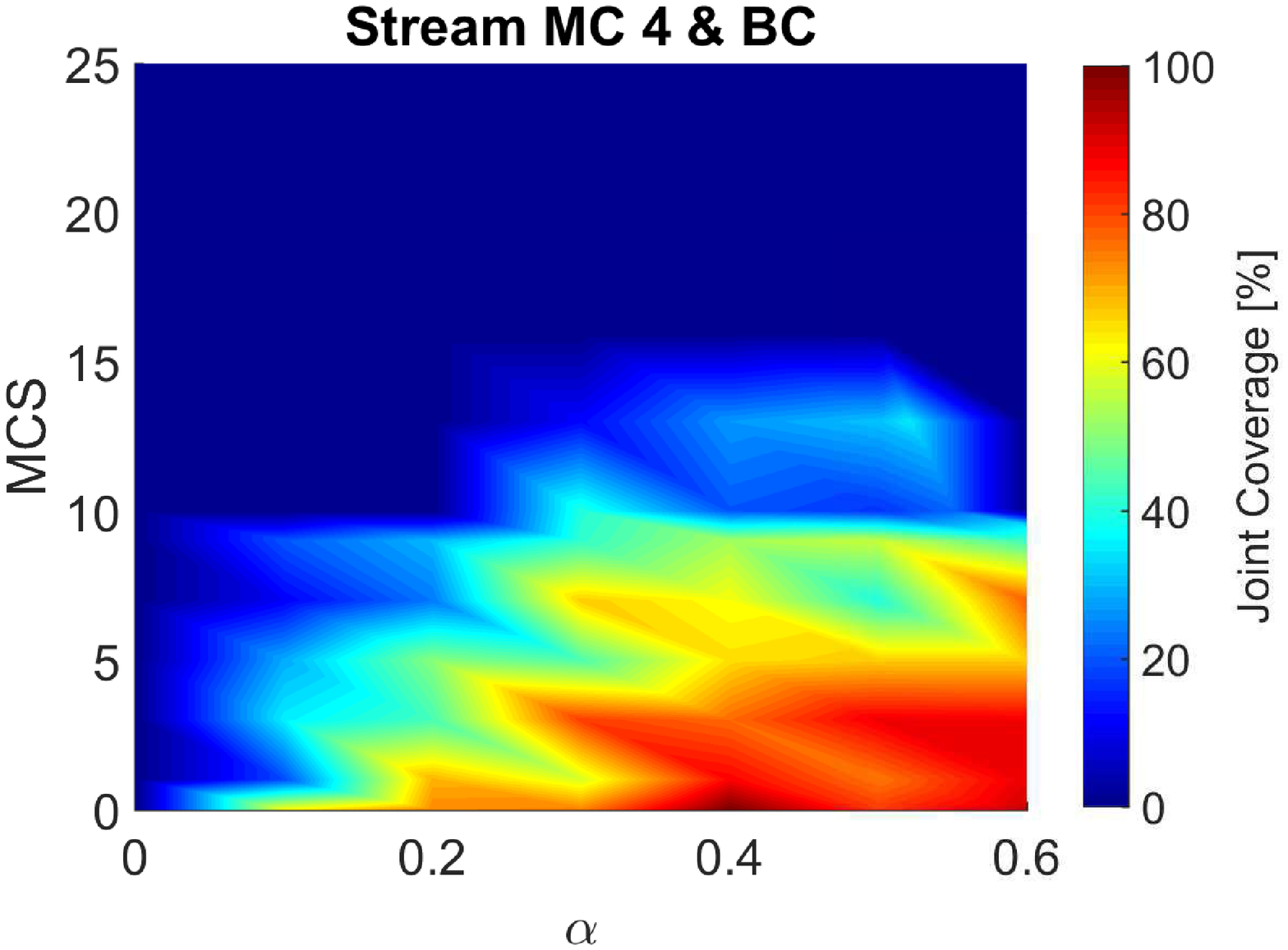}}\hfill
	\caption{Joint Coverage for the SIC Receiver.}
	\label{fig:JointCoverageJD}
\end{figure}

This receiver implements the JD strategy, with also pre-equalization stage. Fig. \ref{fig:JointCoverageJD} depicts the joint coverage metric of each group as a function of MCS and $\alpha$. As in the previous scheme, the maximum joint coverage is achieved for low MCS and for an $\alpha$ in the vicinity of $0.5$. However, the coverage is reduced when it is compared with SIC scheme. 

\section{Implementation and Proof of Concept}
In this section we describe the implementation of the Proof of Concept (PoC) of Joint Broadcast and Multicast Communications. Particularly, we demonstrate the viability of the JD receiver. For this PoC we use $3$ Universal Serial Radio Peripherals (USRP) connected to a General Purpose Processing Unit (GPPU). USRP process the digital samples generated by the GPPU according the framework that we describe in the previous sections, convert the digital signal to/from analog domain and up(down)convert the analog signal to/from the RF band. 

In this PoC one USRP is configured as eNB and other two USRP as UEs. All USRP are connected to the GPPU via $1$ Gigabit Ethernet. All USRP are configured with the same parameters. The cell search stage is performed by using a single carrier frequency, off-line configured. Although the carrier frequency is pre-configured in this setup, UEs shall to synchronize at subframe level. It implies that UEs have to use the synchronization signals provided by LTE standard to determine the Time Advance (TA). Once this synchronization is stablished, UEs have to search the Master Information Block, which carries information about the bandwidth and the number of antennas at eNB. All configuration parameters are estimated by the UEs, except the carrier frequency and the sampling rate. Table \ref{tab:poccom} describes the main configuration parameters used for the demo.

\begin{table}[!ht]
	\centering
	\caption{Common Radio Parameters}
	\begin{tabular}{|>{\bf}c|c|}
		\hline
		Frame Type	& FDD\\\hline
		Prefix Type	& Normal\\\hline
		Carrier frequency &	$915$ MHz\\\hline
		Bandwidth & $5$ MHz\\\hline
		Sampling frequency rate & $7.68$ MS/s\\\hline
		Number of subcarriers & $300$\\\hline
		Control Channel width & $3$ symbols\\\hline
		Coding Rate BC 	& $0.3$\\\hline
		Constellation BC & QPSK\\\hline
		Coding Rate MC1	& $0.43$\\\hline
		Constellation MC1 & QPSK\\\hline
		Coding Rate MC2 & $0.43$\\\hline
		Constellation MC2 & QPSK\\\hline
	\end{tabular}
	\label{tab:poccom}
\end{table}

\begin{figure}[!ht]
	\centering
	\subfloat[eNB]{\includegraphics[width=0.49\linewidth,clip=true]{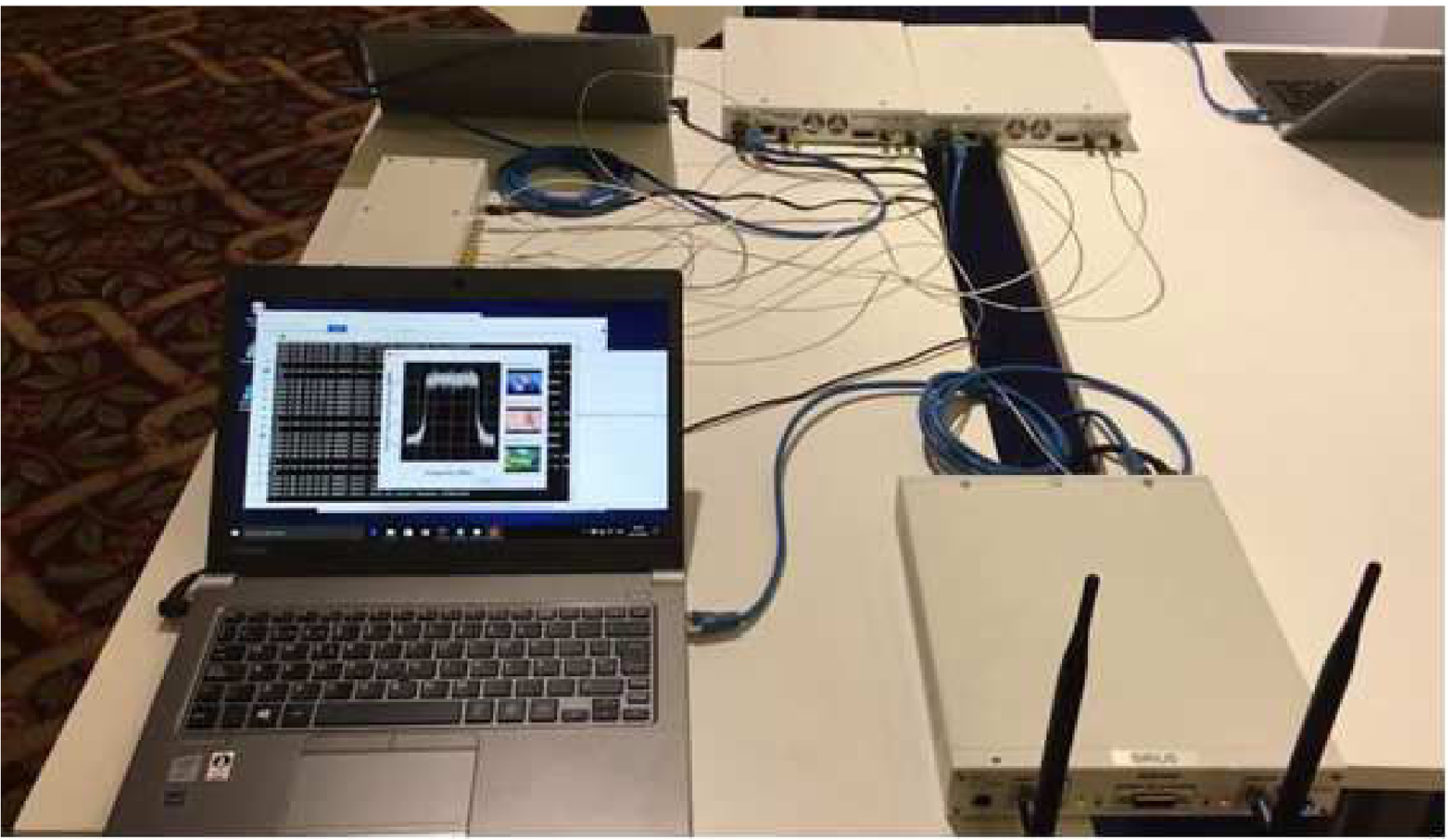}}
	\subfloat[UE]{\includegraphics[width=0.49\linewidth,clip=true]{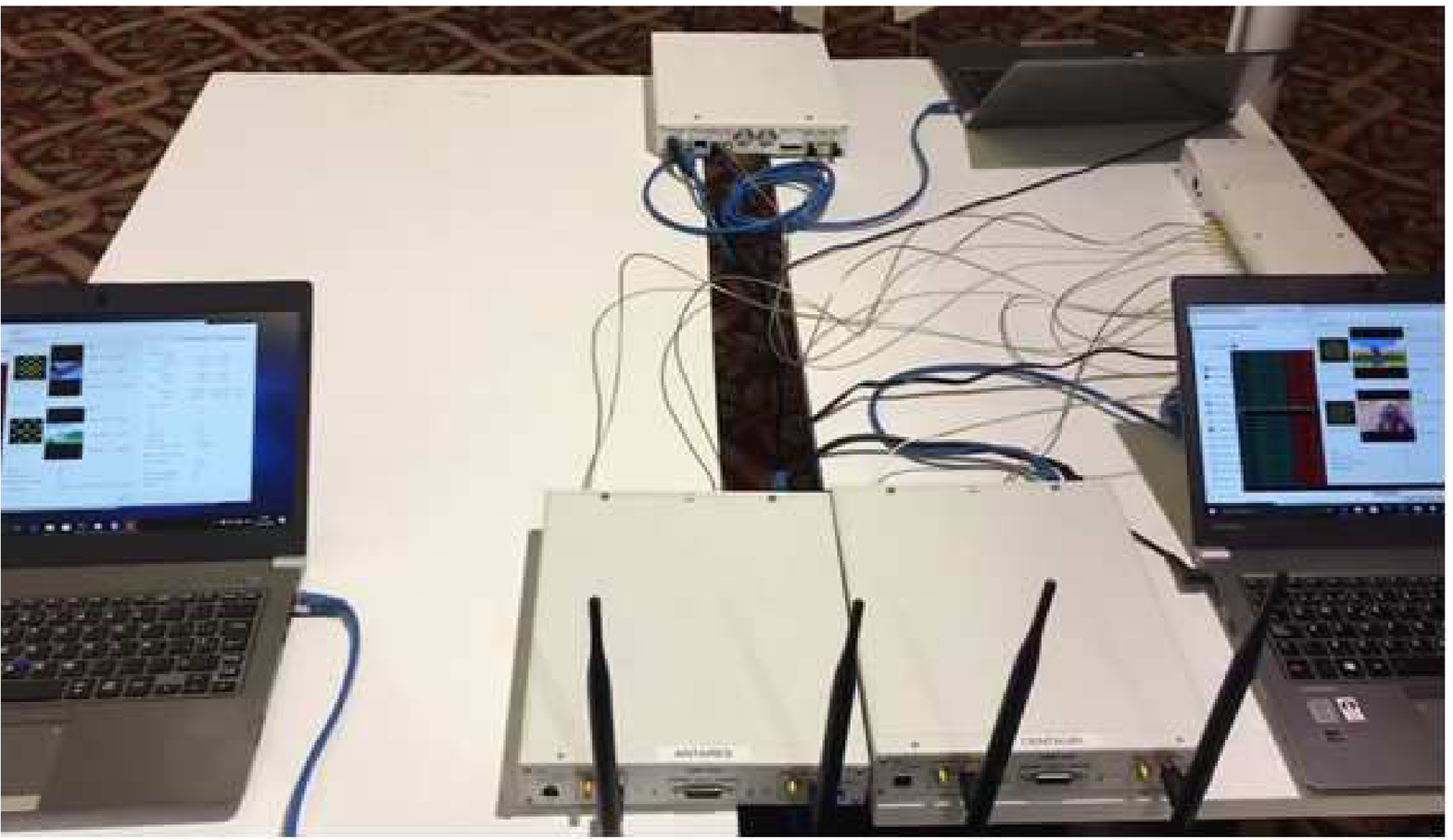}}
	\caption{General overview from eNB and UE sides.}
	\label{fig:PoC}
\end{figure}

Fig. \ref{fig:PoC} contains pictures taken from the PoC where transmitter and receivers are working in real time during the shot. It is important to remark that each receiver decodes the common broadcast video (the same for both receivers) as well as the intended multicast video (different for each receiver). In this PoC, $5$ MHz of bandwidth are shared by the three videos simultaneously.

Constellations reveal which is the impact of superposing streams. Although all streams are conveyed using a QPSK constellation, the superposing introduces cross interferences between the streams and the constellation contains more points compared with the single case. Hence, the receiver has to cancel this interference and recover both streams. It is shown that the scheme is applicable and the demo performance is satisfactory.

\section{Conclusions}
In this paper, we considered the joint broadcast and multicast streaming by allowing non-orthogonal and superposed transmissions. We first introduce the transmission scheme, which thanks to beamforming it is possible to transmit different multicast streams separated by spatial regions in addition to a common superposed broadcast stream, which is received by all users. Different receiving strategies are also tackled. In particular, SIC receiver is employed to reduce the inter-beam interference. A novel Joint Decoder scheme is introduced, which decodes the intended symbols without recursion. Finally, we built a Proof of Concept of the proposed technique, demonstrating its viability. We remark that thanks to the proposed technique, different multimedia services can be delivered without requiring additional bandwidth nor multiplexing in time slots. Sharing the same frequency-resources and allowing different Quality of Service (QoS) enhances the flexibility and efficiency of the system compared with traditional deployments such as LTE.

\section*{Acknowledgment}
This work is supported by the Horizon 2020 project FANTASTIC-5G (ICT-671660), which is partly funded by the European Union, and also supported by the Spanish Government under project ELISA (TEC2014-59255-C3-1-R).

\printbibliography

\end{document}